\documentclass[prd , aps,12 pt,nofootinbib, preprintnumbers,notitlepage]{revtex4-2}
\pdfoutput=1
\usepackage{graphicx}
\usepackage{amssymb}
\usepackage{mathtools}
\usepackage{epstopdf}
\usepackage{color}
\usepackage{bbold}
\usepackage{float}
\usepackage[title]{appendix}
\usepackage{amsfonts}
\usepackage{amsmath}
\usepackage{setspace}
\usepackage{epsfig}
\usepackage[usenames, dvipsnames]{xcolor}
\usepackage{tabularx}
\usepackage{empheq}
\usepackage{mathrsfs}
\usepackage{xcolor}
\usepackage{ulem}
\usepackage[caption=false]{subfig}

\usepackage{units}

\newcommand{\be}{\begin{eqnarray}}
\newcommand{\ee}{\end{eqnarray}}
\newcommand{\beq}{\begin{equation}}
\newcommand{\eeq}{\end{equation}}

\definecolor{bluDT}{cmyk}{1,0.5,0,0.3}
\usepackage[colorlinks, citecolor=bluDT, linkcolor=black, urlcolor=bluDT]{hyperref}

\newcommand{\Ne}{\mathcal{N}_e}

\usepackage{soul}

\begin{document}

\preprint{CERN-TH-2021-213}

\title{\color{bluDT} \LARGE Islands and the de Sitter entropy bound\\
}

\author{\bigskip Daniele Teresi\bigskip}
\affiliation{CERN, Theoretical Physics Department, 1211 Geneva 23, Switzerland\bigskip\bigskip}


\begin{abstract}
The de Sitter (dS) entropy bound gives the maximal number of $e$-folds that non-eternal inflation can last before violating the thermodynamical interpretation of dS space. This semiclassical argument is the analogue, for dS space, of the Black-Hole information paradox. We use techniques developed to address the latter, namely the island formula, to calculate semiclassically the fine-grained entropy as seen by a Minkowskian observer after inflation and find that this follows a Page-like curve, never exceeding the thermodynamic dS entropy. This calculation, performed for a CFT in 2D gravity, suggests that the semiclassical expectation should be modified in such a way that the entropy bound might actually not be present.
\end{abstract}

\maketitle
\onecolumngrid


\section{Introduction} 
The thermodynamical interpretation of de Sitter (dS) space-time is mysterious. One can associate to it a global finite thermodynamic entropy, $S_{\rm dS} \sim M_{\rm Pl}^2/H^2$, although the volume of dS space is infinite. A concrete setup to give an operational meaning to the dS entropy was proposed, some time ago, in~\cite{Arkani-Hamed:2007ryv}. One sees dS space as part of inflation, which globally ends at some time into an asymptotically flat phase. The asymptotic observer is then able to access a number of inflationary modes and associate an entropy to these. When this entropy becomes larger than the thermodynamic entropy one has a paradox. Avoiding this paradox gives rise to the \textit{entropy bound} on the number of $e$-folds of inflation $\Ne \lesssim S_{\rm dS}$.

As pointed out by the authors of~\cite{Arkani-Hamed:2007ryv}, when expressed in terms of this operational setup, this paradox is in some sense analogous to the Black-Hole (BH) information paradox~\cite{Hawking:1976ra}: an observer is able to associate a fine-grained (i.e. quantum-mechanical) entropy to the system, which at some point exceeds its thermodynamic entropy  (i.e. the logarithm of the assumed finite number of available quantum states). In the last couple of years, an enormous progress in the understanding of the BH information paradox has been achieved~\cite{Penington:2019npb,Almheiri:2019psf,Almheiri:2019hni,Penington:2019kki,Almheiri:2019qdq}. For a review, see~\cite{Almheiri:2020cfm}. The lesson learned is that, even if one is in the semiclassical regime, the naive fine-grained entropy of a gravitationally produced system gets modified by the inclusion of \textit{islands}, corresponding to saddles given by extra cuts (along the islands) in the gravitational path-integral that determines the field-theoretical entropy. The correct fine-grained entropy is given by the conjectured \textit{island formula}~\cite{Ryu:2006bv,Ryu:2006ef,Hubeny:2007xt,Faulkner:2013ana,Engelhardt:2014gca,Almheiri:2019hni}, which has been proven explicitly in some particular cases in the BH context~\cite{Penington:2019kki,Almheiri:2019qdq}. 

In this work, we assume that these techniques can also be used to study dS space-time and in particular the setup of~\cite{Arkani-Hamed:2007ryv}. We toy-model the argument of~\cite{Arkani-Hamed:2007ryv} in the language of 2D dilaton gravity and use the island formula to find a Page-like curve~\cite{Page:1993wv,Page:2013dx} for the entropy reconstructed by the asymptotic observer, under a number of assumptions. Then, this fine-grained entropy never exceeds the thermodynamic dS entropy, in the toy-setup considered here. If this result  could be extrapolated to the physical 4D case, it would suggest that the semiclassical expectations leading to the entropy bound should be modified, so that the entropy bound would actually disappear. 

In the last year, a number of other approaches have been developed to study dS space-time by means of the island formula~\cite{Chen:2020tes,Hartman:2020khs,Aalsma:2021bit,Langhoff:2021uct,Aguilar-Gutierrez:2021bns,Kames-King:2021etp,Shaghoulian:2021cef}. 
The overall picture is that non-pathological islands have been  found only when collapsing regions are present, for instance in the analogue of a dS-Schwarzschild system~\cite{Hartman:2020khs,Chen:2020tes}, or in a toy multiverse with collapsing patches~\cite{Aguilar-Gutierrez:2021bns,Langhoff:2021uct}. However, we do not know whether these results should be attributed to the properties of dS space-time, or rather to the BHs present in these examples. 
In this work, instead, we will find that an island appears for a pure dS cosmology (without BHs), as long as we focus on the operational setup giving 
 the entropy bound of~\cite{Arkani-Hamed:2007ryv}.  

The plan of the paper is the following. After this Introduction, in Sec.~\ref{sec:model} we review the entropy bound of~\cite{Arkani-Hamed:2007ryv} and model their setup in the calculable framework of a CFT in 2D gravity, introducing our assumptions, which allow to recover the results of~\cite{Arkani-Hamed:2007ryv} in this simplified context. Then, in Sec.~\ref{sec:island} we show how these assumptions, when used in the island formula, give rise to a Page-like curve for the fine-grained entropy. Finally, in Sec.~\ref{sec:discussion} we discuss our results and draw our conclusions.


\section{de Sitter entropy bound in 2D gravity} \label{sec:model}
\subsection{The entropy bound} \label{sec:entropybound}

Let us briefly review the dS entropy bound of~\cite{Arkani-Hamed:2007ryv}. We consider dS space-time as regularized by a long period of inflation, followed, for simplicity, by flat-space Minkowski evolution. A Minkowskian observer is able to access a number inflationary modes $\approx e^{S}$, associating operationally a semiclassical entropy $S$ to these, since these are semiclassically independent from each other.  Starting from a single Hubble patch, after $\Ne$ $e$-folds of inflation an exponentially large number of patches $e^{(D-1) \Ne}$ is populated, in $D$ space-time dimensions. After a sufficient long period of Minkowski evolution, the observer is then able to associate operationally an entropy $S \sim \Ne$ to inflation. However, when this becomes larger than the dS entropy $S_{\rm dS}$ (with $S_{\rm dS} \sim M_{\rm Pl}^2/H^2$ in $D=4$), we have a paradox, analogous to the BH information paradox. As a consequence, \cite{Arkani-Hamed:2007ryv} obtains the \textit{entropy bound} $\Ne \lesssim S_{\rm dS}$.

\subsection{CFT entropy in 2D Gravity}

\begin{figure}[t]
$$\includegraphics[height=12em]{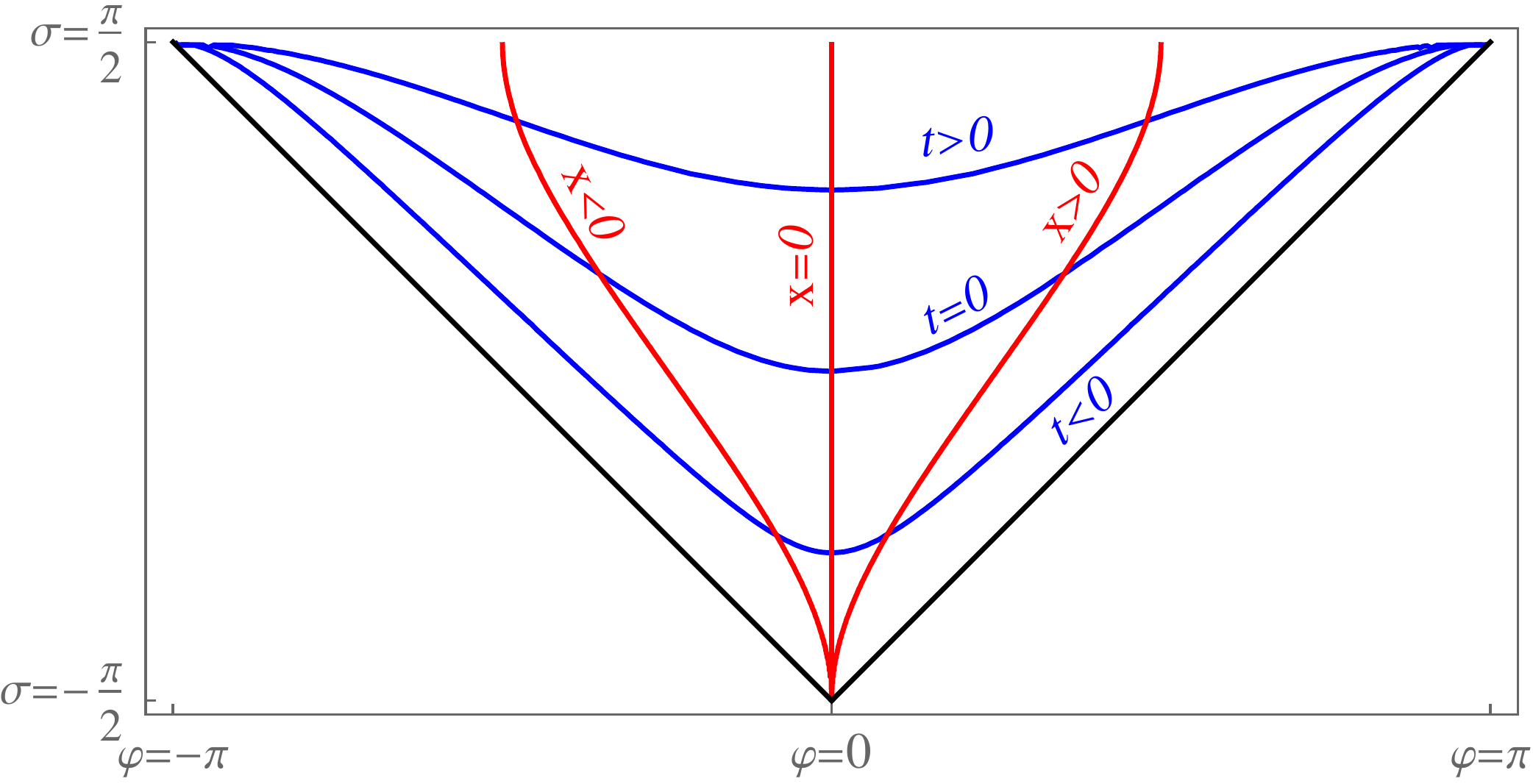}$$
\caption{Conformal diagram of 2D dS space, where both the global coordinates \eqref{eq:global} and the planar ones \eqref{eq:planar} are shown. The latter, usually adopted for cosmic inflation, cover only half of dS space, the inflationary patch. \label{fig:coords}}
\end{figure}

Our aim is to model this setup in a calculable framework, namely a CFT in 2D gravity, as we now describe. While in 1+1 dimensions gravity is purely topological, a dynamical dS theory is obtained by including a dilaton, i.e. the dS Jackiw-Teitelboim action~\cite{Jackiw:1984je,Teitelboim:1983ux,Almheiri:2014cka,Maldacena:2019cbz,Cotler:2019nbi}: 
\beq
S_{JT} = \frac{1}{16 \pi G} \int d^2 x \sqrt{- g} \left [ \phi_0 R + \phi (R-2 H^2)\right] \, , \label{eq:JT}
\eeq
up to boundary terms. The solution describing global dS space (without BHs) in global coordinates is:
\beq 
d s^2 = \frac{1}{H^2 \cos^2\sigma } ( - d \sigma^2 + d \varphi^2) \, , \qquad \phi = \phi_r \tan \sigma \, ,   \label{eq:global}
\eeq
with $-\pi/2 \leq \sigma \leq \pi/2$, and $-\pi \leq \varphi \leq \pi$ with the extrema identified. The conformal diagram is shown in Fig.~\ref{fig:coords}. The dilaton profile is effectively the inverse gravitational coupling, so that, by taking $\phi_r > 0$, gravity becomes strongly coupled in the past $\sigma \to -\pi/2$ and weakly coupled in the future.  For a given relevant range of $\sigma$, we can avoid the strongly-coupled regime $\phi_0 + \phi \to 0$ by choosing $\phi_r$ sufficiently small.  

In order to calculate analytically the entanglement entropy associated to a given region, we include a CFT with a central charge $c \gg 1$, so that its many degrees of freedom dominate the entropy (and in 2D its entanglement entropy is well known), without affecting the dS character of spacetime. The entanglement entropy depends on three ingredients: the quantum state of the CFT, the region under consideration and the UV cutoff selecting the modes included in the analysis. About the first, since we are interested in modelling inflation we consider the Bunch-Davies vacuum. The standard procedure to obtain the entanglement entropy in this state is the following. We first introduce conformal complex coordinates
\beq 
z = e^{-i (\sigma - \varphi)} \,, \qquad \bar{z} = e^{-i (\sigma + \varphi)} \, , \qquad d s^2 = \frac{1}{\Omega(z,\bar z)^2} \, dz d \bar z \, ,
\eeq 
with $\Omega = H (1 + z \bar z)/2$. The vacuum state in this coordinate system is known to be the Bunch-Davies (or Hartle-Hawking) vacuum ($z, \bar z$ are light-cone coordinates). By means of a conformal transformation, we can then use the flat spacetime well-known result~\cite{Calabrese:2004eu}, by keeping track of the conformal transformation of the UV cutoff, and obtain the entropy of a single interval
\beq
S = \frac{c}{6} \log \frac{(z_1 - z_2)(\bar z_1 - \bar z_2)}{\epsilon_1 \epsilon_2 \, \Omega_1 \Omega_2} \,, \label{eq:one_interval}
\eeq
where the subscript  denotes quantities calculated at the corresponding endpoint of the interval and $\epsilon$ is the cutoff expressed in the relevant system of coordinates. Written in global coordinates this becomes
\beq
S = \frac{c}{6} \, \log \frac{2 \, [ \cos (\sigma_2 - \sigma_1) - \cos (\varphi_2 - \varphi_1)]}{\cos \sigma_1 \cos \sigma_2 \, H^2 \epsilon_1 \epsilon_2} \,. \label{eq:Sglobal}
\eeq
In the following, it will be clear that for our purposes it is important to retain the global structure of the Bunch-Davies/Hartle-Hawking vacuum rather than, for instance, considering its late time  (simpler) Poincaré limit.

\subsection{Modelling the entropy bound}
The remaining two ingredients are the region of interest and the UV cutoff of the modes considered. In their choice, we are guided by the aim of modelling, as much as possible, the operational setup giving the entropy bound, described in Sec.~\ref{sec:entropybound}. 

\begin{figure}[t]
$$\includegraphics[height=16em]{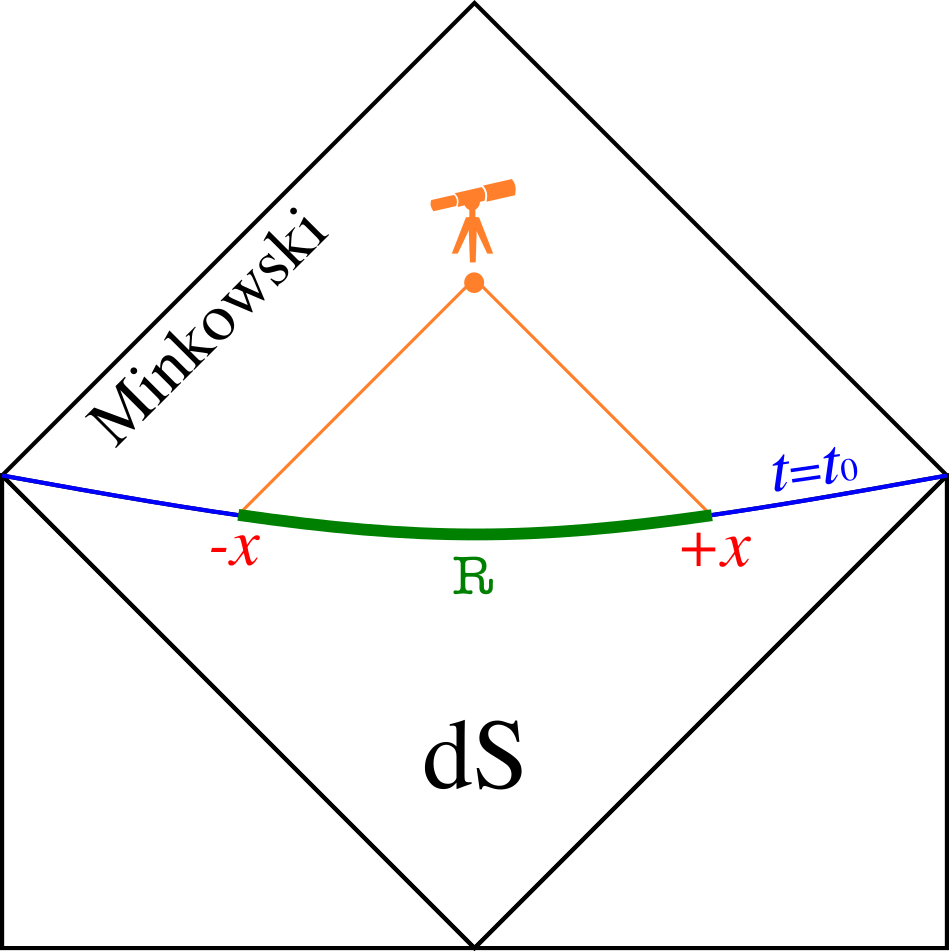}$$
\caption{Sketch of the framework modelling the operational setup of Sec.~\ref{sec:entropybound}. The dS stage ends at $t=t_0$ and Minkowski evolution starts. After some time, a Minkowskian observer (orange dot) is able to observe the inflationary modes in the region $\mathcal{R}$.\label{fig:sketch}}
\end{figure}

We do not have an inflaton. However, in modelling the setup above we may take it as a \textit{clock} and reason in terms of the usual planar coordinates of inflation (shown in Fig.~\ref{fig:coords}):
\beq
d s^2 = - d t^2 + e^{2 H t} d x^2 \, , \qquad e^{H t} = \frac{\cos \varphi + \sin \sigma}{\cos \sigma} \,, \qquad H x = \frac{\sin \varphi}{\cos \varphi + \sin \sigma} \, , \label{eq:planar}
\eeq
with $-\infty < t,x < + \infty$, imagining that dS ends at some fixed time $t_0$, modelling the inflaton value for reheating. Then, Minkowski evolution starts and an observer, after some time, will have access to the CFT inflationary modes at the reheating surface, associating an entropy to these. As a consequence, the entropy appearing in the operational setup of Sec.~\ref{sec:entropybound} is nothing but the fine-grained (entanglement) entropy of a region $\mathcal{R}$ given by an interval $(-x, x)$ at fixed $t = t_0$. This is sketched in Fig.~\ref{fig:sketch}.

The last ingredient is the UV cutoff. While originally the UV cutoff is introduced to regulate the divergence for an interval of vanishing length, we assume that it can be interpreted as the cutoff of the relevant modes, whose entropy one is focused on\footnote{In other words, one can trace over the remaining modes in the UV. For free fields, this is the same as imposing a relevant UV cutoff. For interacting fields, one should consider a Wilsonian effective action.}.  In order to model the setup of Sec.~\ref{sec:entropybound}, we choose a cutoff
\beq
\epsilon \sim \frac{1}{H} \,, \label{eq:cutoff}
\eeq
i.e at $t=t_0$ we consider all and only those modes that have been frozen during inflation. These are precisely the ones that the future Minkowskian observer can reconstruct, for instance from the CMB, in the argument of Sec.~\ref{sec:entropybound}.
 We anticipate that in the next Section we will argue that this prescription is suitable for the region $\mathcal{R}$, but not for the island in the gravitating dS region. 

Having specified all the relevant ingredients, we can now calculate the semiclassical entropy seen by the future observer, by combining \eqref{eq:Sglobal}, \eqref{eq:planar} and \eqref{eq:cutoff}, and obtain the simple result
\beq
S_{\rm semi} \simeq \frac{c}{3} \log \frac{2 e^{H t_0} x}{\epsilon} \sim \frac{c}{3} \log (H l_\mathcal{R})  \;, \label{eq:Ssemi}
\eeq
having introduced $l_\mathcal{R} = 2 e^{H t_0} x$, the proper length of $\mathcal{R}$. 
The entropy is simply counting the modes that exited the horizon during inflation.

Finally, let us see how the argument of Sec.~\ref{sec:entropybound} is reproduced. Let us analogously assume that inflation starts at some time $t_i < 0$. Starting from a single Hubble patch, inflation fills a comoving distance $x \sim e^{- H t_i}/H$. Then, \eqref{eq:Ssemi} and \eqref{eq:cutoff} give that the maximal semiclassical entropy that the Minkowskian observer is able to reconstruct (sufficiently in the future) is:
\beq
S_{\rm semi} \sim \frac{c}{3} \log \frac{e^{H (t_0 - t_i)}}{H \epsilon} \sim \frac{c}{3} \log e^{H (t_0 - t_i)} \sim \Ne \,, \label{eq:Ssemitot}
\eeq
as in Sec.~\ref{sec:entropybound}. When this fine-grained entropy becomes larger than the thermodynamic dS entropy\footnote{This is the horizon area in Planck units. In 2D, $\phi_0$ plays the role of the Planck mass and the factor of 2 comes from the horizon being two points.} $S_{\rm dS} \sim 2 \phi_0$ we have an apparent paradox, since the latter measures the total number of states available to the quantum system that should be describing dS space\footnote{Assuming the analogue of the \textit{central dogma} for BHs, i.e. that the meaning of the dS entropy is indeed thermodynamical. If the dS entropy had some other unknown meaning, the entropy bound would not be present to start with.}.

\section{The entropy after a long period of inflation} \label{sec:island}

The enormous progress of the last few years in the understanding of the BH information paradox~\cite{Penington:2019npb,Almheiri:2019psf,Almheiri:2019hni,Penington:2019kki,Almheiri:2019qdq,Almheiri:2020cfm} is based on the appreciation that for a system produced gravitationally a semiclassical entropy such as~\eqref{eq:Ssemi} fails to give the correct result, at late times. Instead, the correct fine-grained entropy is given by the island formula~\cite{Ryu:2006bv,Ryu:2006ef,Hubeny:2007xt,Faulkner:2013ana,Engelhardt:2014gca,Almheiri:2019hni}:
\beq
S(\mathcal R) = \min \mathrm{ext}_{\mathcal I} \left ( \frac{\text{Area}(\partial \mathcal I) }{4} + S_{\rm semi}(\mathcal R \cup \mathcal I)\right) \, , \label{eq:island}
\eeq
where the island $\mathcal{I}$ is in the gravitating region and the minimum among the different extrema (including $\mathcal I = \varnothing$) is taken. The area of the island boundary $\partial \mathcal{I}$ is in Planck units. Its explicit proof for eternal BHs in 2D gravity~\cite{Penington:2019kki,Almheiri:2019qdq} suggests the following interpretation. In QFT the entanglement entropy is calculated in terms of Euclidean path integrals on appropriate replica manifolds, where the different sheets are connected by cuts in the region of interest $\mathcal{R}$. In general, since in the gravitational path integral spacetime itself is dynamical, the dominant configurations may contain extra cuts on $\mathcal{I}$, if gravity is allowed to ``decide'' the replica topology, as dominant saddles in the Euclidean gravitational path-intergral. The extra cuts introduce conical singularities at their boundary, captured by the area term in~\eqref{eq:island}, which gives the entropic contribution of gravity itself.

\begin{figure}[t]
$$\includegraphics[height=12em]{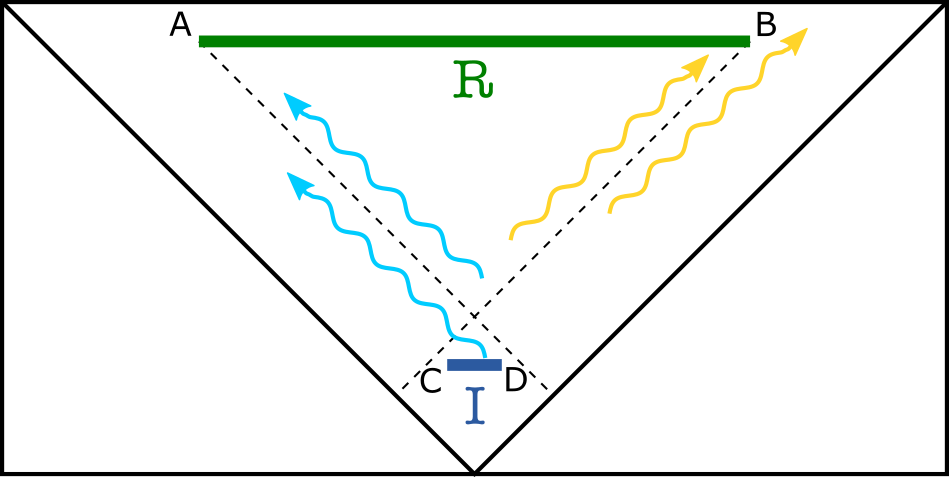}$$
\caption{Island contribution to the entropy. The island $\mathcal{I}$ is shown as a cut joining the two endpoints from ``inside''.
We  show modes that stop contributing to the entropy of $\mathcal{R}$ when the island $\mathcal{I}$ is included. Pairs with the same colour are entangled.
\label{fig:sketchisland}}
\end{figure}

Therefore, we look for an island in the setup considered so far. Since we do not know a priori if we will be in a simplifying universal limit (where the CFT entropy is simplified by an OPE), we consider the two-intervals entropy formula. This is known only for a specific CFT, namely $c$ free fermions, so that we now restrict to this case. As before, we can do a conformal transformation from conformal complex coordinates and use the flat-space vacuum entropy~\cite{Casini:2005rm} to obtain
\beq
S = \frac{c}{6} \log \frac{\ell^2_{AB} \ell^2_{BC} \ell^2_{CD} \ell^2_{DA}}{\ell^2_{AC} \ell^2_{BD} \epsilon_A \epsilon_B \epsilon_C \epsilon_D \Omega_A \Omega_B \Omega_C \Omega_D} \,, \label{eq:two_intervals}
\eeq
where the different letters label the endpoints of $\mathcal R$ and $\mathcal I$ as denoted in Fig.~\ref{fig:sketchisland} and we have introduced the short-hand notation $\ell^2_{XY} \equiv (z_X - z_Y)(\bar z_X - \bar z_Y)$.

Notice that the points $C$ and $D$ (as well as $A$ and $B$) are differentiated by the different direction of wrapping around them to move to the same replica sheet. In particular, a counter-clockwise  wrapping of $2 \pi$ around $A$ and $C$ shifts the replica sheet as $n \to n+1$, whereas as $n \to n-1$ for $B$ and $D$. As a consequence, swapping $C$ and $D$ while keeping $A$ and $B$ fixed changes the replica manifold and hence the entropy\footnote{For similar reasons, one can equivalently see the cut as shifting $n \to n+1$ and connecting $C$ and $D$ from ``inside'', or as shifting $n \to n-1$ and connecting them from ``outside''. The complementary choices, instead, give a different replica manifold.}.

From \eqref{eq:two_intervals} and Fig.~\ref{fig:sketchisland} we may already understand why an island, if appearing, can compete with the semiclassical entropy~\eqref{eq:Ssemi} beating the penalty for the area term in~\eqref{eq:island}. In the limit in which $BC$ and $AD$ become light-like their contribution suppresses the numerator of~\eqref{eq:two_intervals} and compensates for the large contribution of $AB$ (which essentially gives the semiclassical entropy~\eqref{eq:Ssemi}). Moreover, the area term
\beq
\frac{\text{Area}(\partial \mathcal I) }{4} = 2 \phi_0 + 2 \phi_r \tan \sigma_{\mathcal{I}}
\eeq
becomes smaller and smaller going towards the past. Here we have set $4G=1$ (which is a normalization for $\phi_0$) and taken $\sigma_C = \sigma_D \equiv \sigma_{\mathcal{I}}$. Below, we will find that the OPE approximation is not appropriate for the determination of the dominant island, since this appears only when the full structure in \eqref{eq:two_intervals} is retained. Therefore, we cannot know whether our results extrapolate to a generic CFT or not.

While the cutoffs $\epsilon_A, \epsilon_B$ in \eqref{eq:two_intervals} are in the region $\mathcal{R}$ and can  be hence chosen at will, e.g. as given by~\eqref{eq:cutoff}, the situation is different for the the cutoffs $\epsilon_C, \epsilon_D$ in the island region\footnote{We thank Victor Gorbenko for pointing  this out to us.}. This is because even if we focus on a subset of all modes (the ones that have been frozen during inflation), the emergent cut in the replica manifold given by the island affects all modes in the theory, even if they are traced over in $\mathcal{R}$. Therefore, the inclusion of this cut for them is equivalent to consider the CFT entropy as cut off by the real UV cutoff of the theory $\epsilon_{\rm UV}$ (presumably a proper length, because of covariance). Then, the standard lore (see e.g.~\cite{Bousso:2015mna,Hartman:2020khs}) is that the UV divergence is absorbed into the renormalization of the inverse gravitational coupling $\phi_0$, so that the full result~\eqref{eq:island} is finite. Effectively, one replaces $\epsilon_{\rm UV} \to \epsilon_{\rm RG}$, the renormalization scale that we leave unspecified.

Taking this into account, the two-intervals entropy \eqref{eq:two_intervals} in global coordinates becomes
\beq
S_{\rm semi} = \frac{c}{6} \log \frac{4 \, (1 - \cos 2 \varphi_\mathcal{R})(1 - \cos 2 \varphi_\mathcal{I})}{H^4 \epsilon_{\rm RG}^2 \epsilon^2  \cos^2 \sigma_\mathcal{R} \cos^2 \sigma_\mathcal{I}} \frac{[\cos (\sigma_\mathcal{R} - \sigma_\mathcal{I}) - \cos (\varphi_\mathcal{R} + \varphi_\mathcal{I})]^2}{[\cos (\sigma_\mathcal{R} - \sigma_\mathcal{I}) - \cos (\varphi_\mathcal{R} - \varphi_\mathcal{I})]^2} \,, \label{eq:Ssemi_island}
\eeq
where we have already used symmetry to impose $\sigma_A = \sigma_B \equiv \sigma_{\mathcal{R}}$, $\varphi_B = - \varphi_A \equiv \varphi_{\mathcal{R}} > 0$,  $\varphi_D = - \varphi_C \equiv \varphi_{\mathcal{I}}$. In view of the discussion above after~\eqref{eq:two_intervals}, positive and negative values of $\varphi_\mathcal{I}$ correspond to different replica manifolds, where $C$ and $D$ are swapped.

\begin{figure}[t]
$$\includegraphics[height=14em]{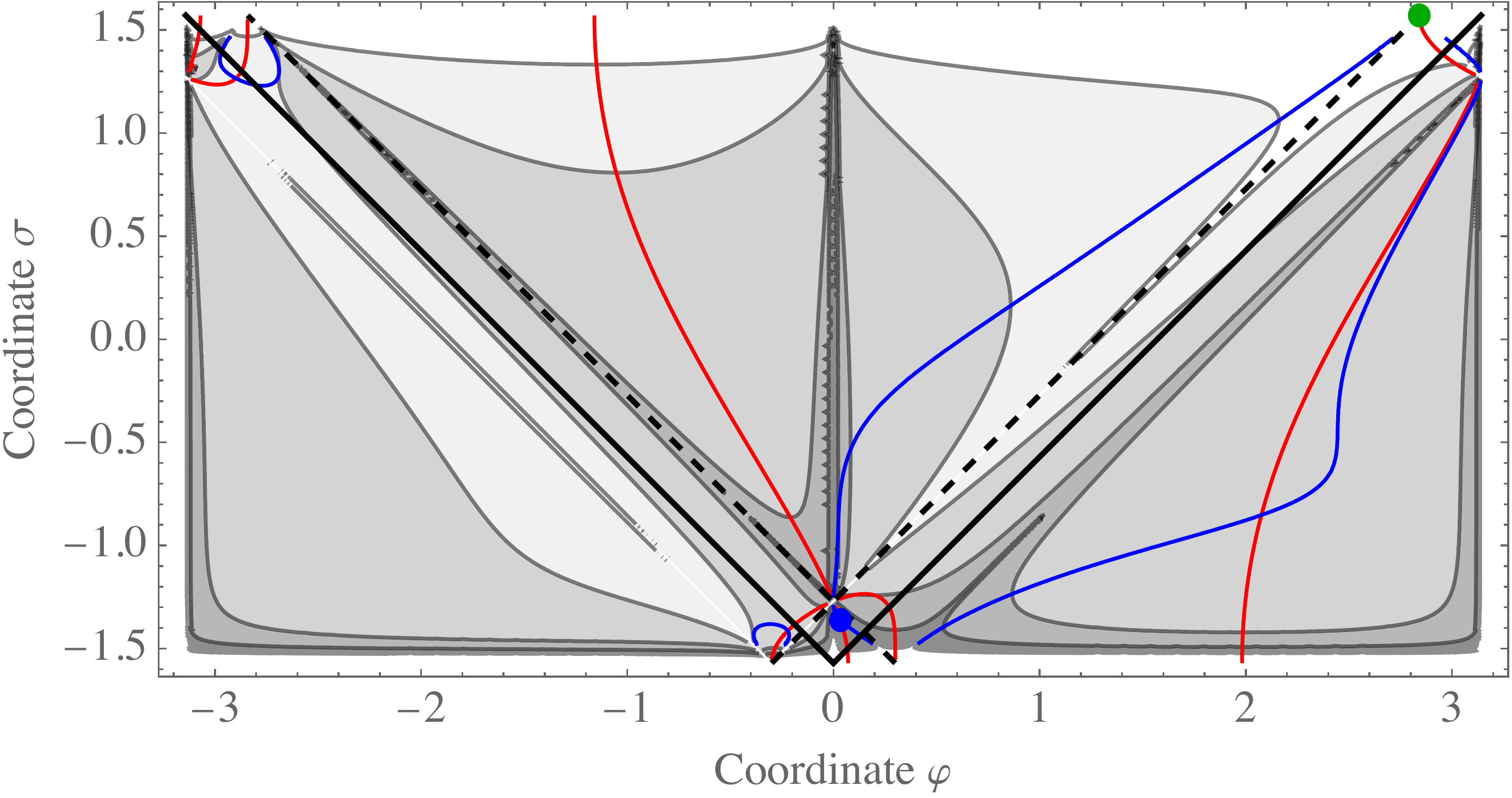}$$
\caption{Isocontours of the island entropy $S$ as given in~\eqref{eq:S_island} (shaded in lighter gray for larger $S$) and contours of vanishing partial derivatives with respect to $\sigma_I$ (blue lines) and $\varphi_I$ (red lines). The green dot denotes the endpoint $B$ of the region $\mathcal{R}$, at large times $t_0 \to \infty$, i.e. $\sigma_\mathcal{R} = \pi/2$, while the blue dot denotes the island endpoint $D$. The dashed lines indicate the past lightcone of $\mathcal{R}$.\label{fig:global}}
\end{figure}

\begin{figure}[t]
\centering
\subfloat[\label{fig:close_past}]{\includegraphics[height=12em]{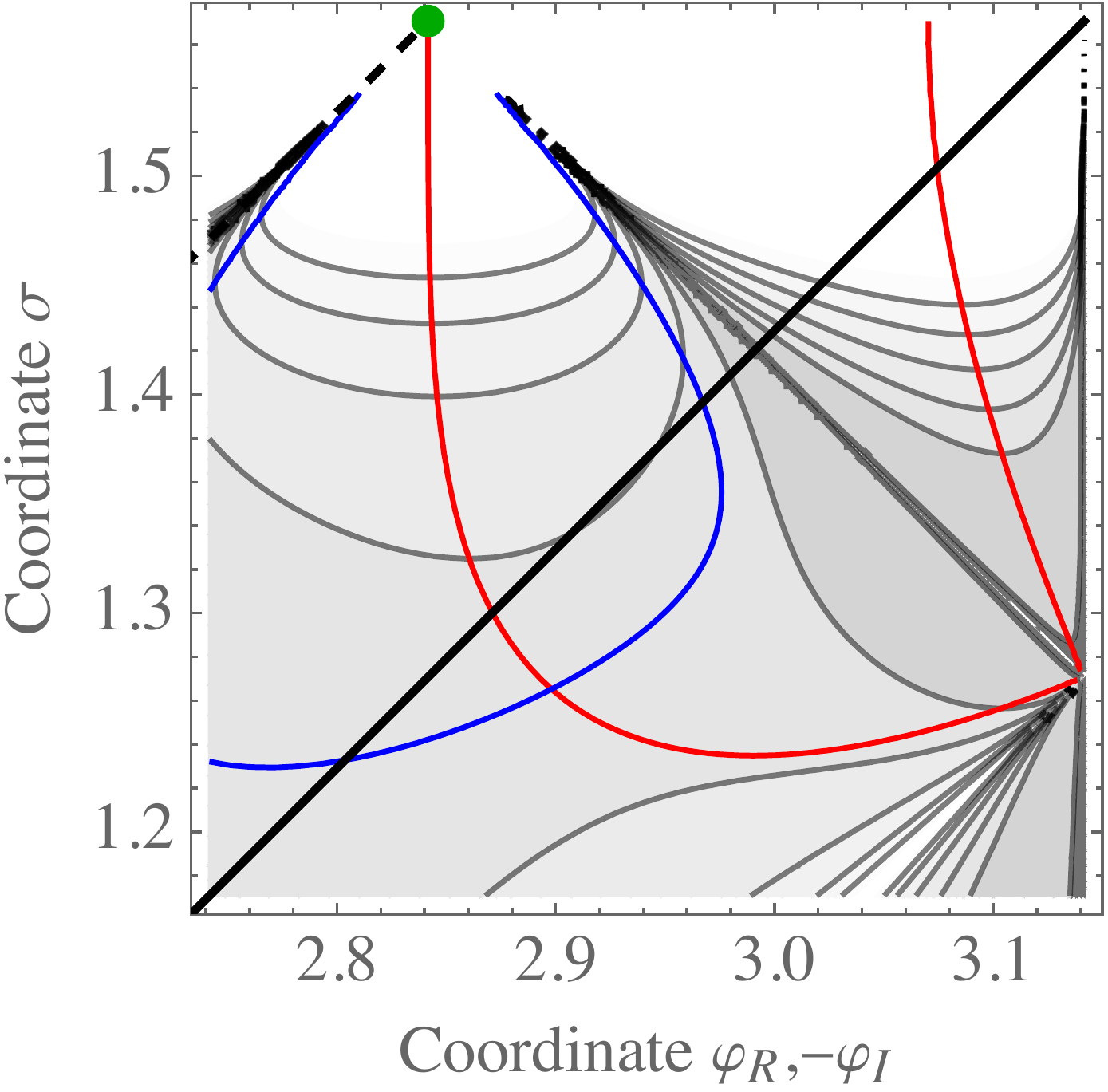}} \qquad
\subfloat[\label{fig:far_past}]{\includegraphics[height=12em]{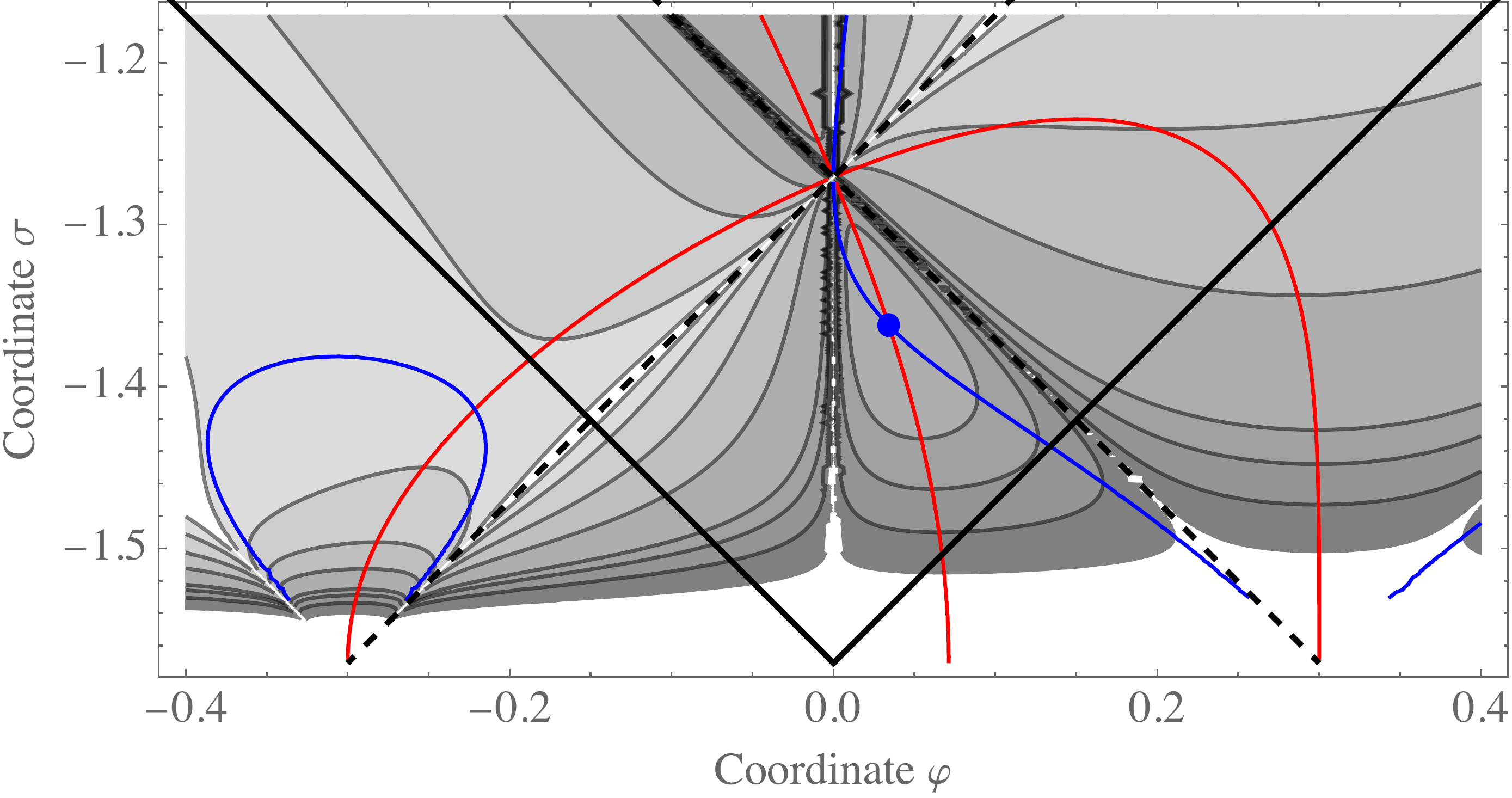}}
\caption{Zoom of Fig.~\ref{fig:global} in the regions in the close  and distant past (left and right panels, respectively). Extrema of $S$ arise at the intersection of blue and red lines. In addition to the island, denoted by the blue dot, other extrema are also present, but the corresponding entropy is larger (lighter shading). In the left panel, for convenience we have swapped the sign of the island coordinate $\varphi_{\mathcal I} \to -\varphi_{\mathcal I}$, see text. \label{fig:local}}
\end{figure}

The total entropy in the presence of the island is 
\beq
S(\sigma_\mathcal{I}, \varphi_{\mathcal{I}}) = 2 \phi_0 + 2 \phi_r \tan \sigma_{\mathcal{I}} + S_{\rm semi} \,, \label{eq:S_island}
\eeq
with $S_{\rm semi}$ as given in~\eqref{eq:Ssemi_island}. We need to find the extrema of this quantity. This is best done numerically, as shown in Figs.~\ref{fig:global} and~\ref{fig:local} for $t_0 \to \infty$. There are several extrema. In the close past (Fig.~\ref{fig:close_past}) there is a minimax (i.e. minimal in time, maximal in space) saddle-point. This is analogous to the island found in~\cite{Chen:2020tes} in a similar setup, which is argued to be pathological, since it violates the strong-sub-additivity condition by a large amount\footnote{The minimax saddle-point found in~\cite{Shaghoulian:2021cef} in a different setup is also argued to be pathological.}. The fact that $\varphi_{\mathcal{I}}$ is negative just means that the direction of wrapping around the island endpoints is swapped, as explained above after~\eqref{eq:Ssemi_island}. However, we find that the dominant extremum (the one with minimal $S$, as prescribed by the island formula) is in the distant past, as shown in Fig.~\ref{fig:far_past}. This is a local maximum of $S$. While the former extremum is captured by the OPE approximation and the Poincaré limit of the vacuum, the island $\mathcal{I}$ is present only when the full structure~\eqref{eq:two_intervals} is retained. 

\begin{figure}[t]
\centering 
\includegraphics[height=12em]{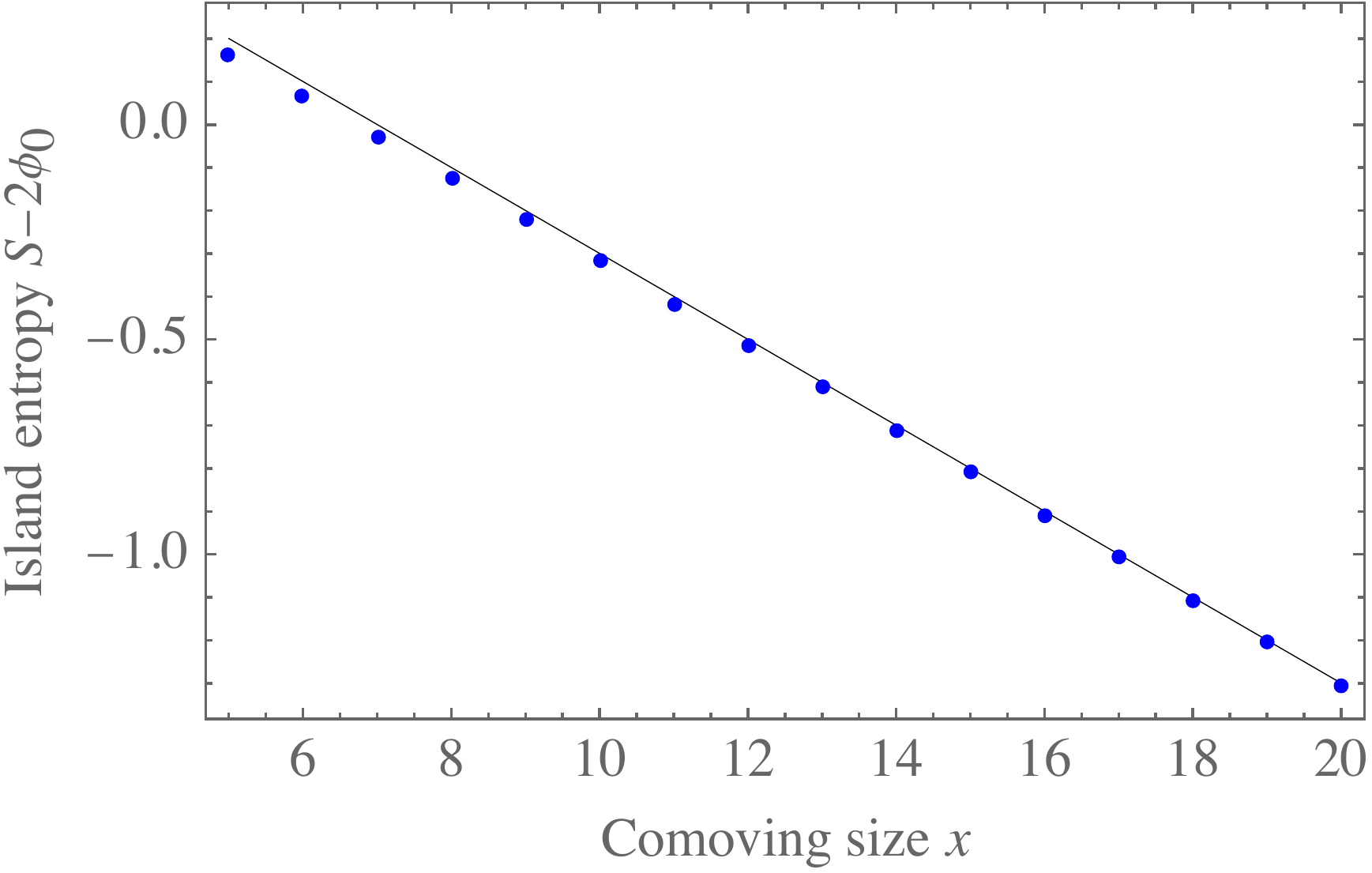}
\caption{Analytic approximation to the island entropy (continuous line), compared to the full numerical results (blue dots) for $t_0 \to \infty$. We worked in Hubble units and chose the example values of the parameters $c=1$, $\phi_r=0.1$, $\varepsilon=0.1$, $\epsilon_{\rm RG}=1$. \label{fig:numerics}}
\end{figure}

Having established the existence of the island, we can approximate its entropy analytically for $t_0 \to \infty$ and
$\delta \equiv \pi - \varphi_\mathcal{R} \simeq 2/ H x \to 0$.
This is done by expanding $\partial S/\partial \varphi_I = \partial S/\partial \sigma_I = 0$  for  small $\varphi_\mathcal{I}$ and $\sigma_\mathcal{I} + \pi/2$ and finding the approximate location of the island:
 \beq
 \varphi_{\mathcal{I}} \simeq \frac{c}{12 \phi_r} \delta^2  \,, \qquad \sigma_{\mathcal{I}} + \frac{\pi}{2} \simeq \delta - \frac{c}{6 \phi_r} \delta^2 \,. 
 \eeq
Then, we can restore a finite $t_0$, i.e.~a finite $\pi/2-\sigma_\mathcal{R}$, only in the factor $\cos \sigma_\mathcal{R}$ in \eqref{eq:Ssemi_island}, which is formally singular in the $t_0 \to \infty$ limit.
We thus obtain the approximate form of the fine-grained entropy:
\beq\label{eq:Page}
S(\mathcal{R}) \simeq \min \bigg\{\frac{c}{3} \log \frac{2 e^{H t_0} x}{\epsilon} , \; 2\phi_0 - \phi_r H x + \frac{c}{3}  \bigg(\log \frac{2 e^{H t_0} c}{9 \phi_r H^2\epsilon_{\rm RG}  \epsilon } - 1\bigg)\bigg\} \,,
\eeq
having included the no-island contribution. In Fig.~\ref{fig:numerics}, we compare the analytical approximation of the island contribution to the full numerical result, showing that the former is indeed sufficiently accurate\footnote{Notice however that the $\mathcal{O}$(1) factors inside the log in~\eqref{eq:Ssemi} and \eqref{eq:Ssemi_island} (and hence in~\eqref{eq:Page}) should not be taken seriously, since the formulas used for the CFT entropies capture only the log behaviour.}. The fine-grained entropy can be expressed directly in terms of physical quantities\footnote{The original explicit dependence on $t_0$ is fictitious. Otherwise this would be unphysical, because the value of $t_0$ depends on the chosen origin of planar time. The key point is that $\phi_r$ implicitly depends on $t_0$: the dilaton profile at $t_0$, which gives the physical gravitational coupling at the end of the dS phase, is asymptotically given by $\phi \simeq \phi_0 + \phi_r e^{H t_0} x^2/2 = \phi_0 + \phi_r e^{-H t_0} l_\mathcal{R}^2/2 $, in terms of the physical length $l_{\mathcal R}$. Therefore, the physical combination is $\phi_r/e^{H t_0} \equiv \phi_r^{\rm ph}$.} $l_\mathcal{R}$, $\phi_r^{\rm ph} \equiv \phi_r/e^{H t_0}$ as
\beq\label{eq:Page_ph}
S(\mathcal{R}) \simeq \min \bigg\{\frac{c}{3} \log (H l_\mathcal{R}), \; 2\phi_0 - \frac{\phi_r^{\rm ph} H l_\mathcal{R}}{2} + \frac{c}{3}  \bigg(\log \frac{2 c}{9 \phi_r^{\rm ph} H \epsilon_{\rm RG}} - 1\bigg)\bigg\} \,.
\eeq

The entropy of $\mathcal{R}$ follows a Page-like curve, as also shown in Fig.~\ref{fig:Page}. As a consequence, the Minkowskian future observer of Sec.~\ref{sec:entropybound} is never able to measure an entropy larger than the thermodynamic one and the paradox disappears. This suggests that the semiclassical expectation should be modified in such a way that the entropy bound  might actually not be present.

\section{Discussion and conclusions} \label{sec:discussion}
To summarize, the basic assumptions that lead to the Page-like curve for the entropy, as reconstructed by the Minkowskian observer, are the following
\begin{enumerate}
\item The operational setup that gives the entropy bound can be (toy-)modelled by 2D gravity with a CFT with large central charge.
\item The island formula gives the correct fine-grained entropy in the dS setup considered, in particular for an island not on the same Cauchy slice as the region of interest.
\item In this context a local maximum is a legitimate extremum for the island formula.
\end{enumerate}
Under these assumptions, we have found that the fine-grained entropy is given by~\eqref{eq:Page} and, in particular, never exceeds the thermodynamic dS entropy $S_{\rm dS} \sim 2 \phi_0$. 

In detail, we found that the island contribution dominates for comoving sizes larger than $x_*$, with
\beq
H x_* \simeq \frac{2 \phi_0}{\phi_r} - \frac{c}{3 \phi_r} \log \frac{2 e^{H t_0} x_*}{\epsilon } \,,
\eeq
and the maximal entropy
 is $S_* = S(x_*) \approx \frac{c}{3} \log(4 \phi_0/\phi_r^{\rm ph})$.

\begin{figure}[t]
$$\includegraphics[height=14em]{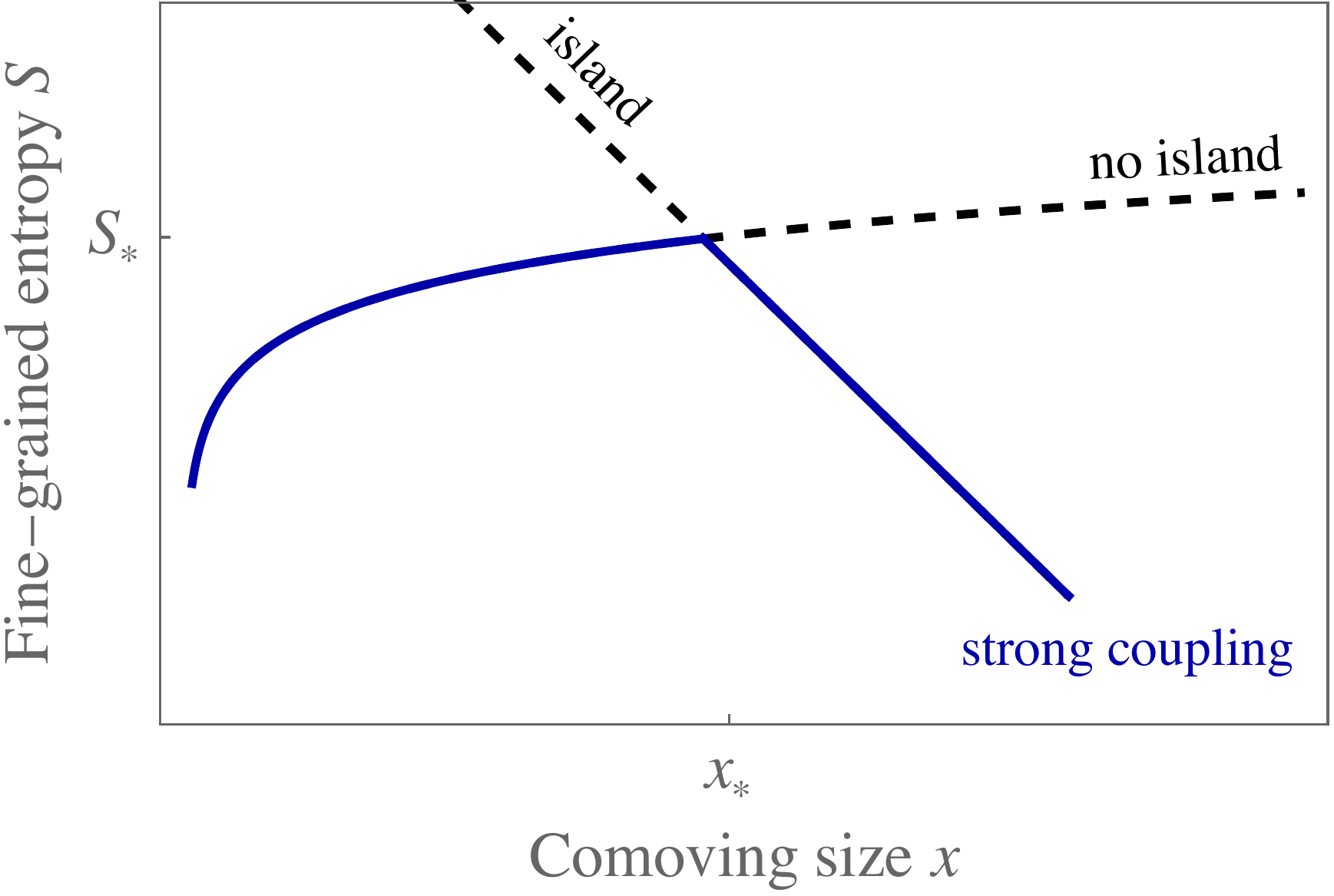}$$
\caption{Page-like curve for the fine-grained entropy, as function of the comoving size $2x$ of the region $\mathcal{R}$, at $t_0 \to \infty$. \label{fig:Page}}
\end{figure}

Notice that $S_*$ is generally much smaller than $S_{\rm dS}$. However, this might be a peculiarity of the 2D toy-model, since it is ultimately due to the fact that both the past light-cone and horizon areas go to zero (in Planck units) in the far past, for the theory~\eqref{eq:JT} (see also Fig.~\ref{fig:sketchisland}). Instead, for 4D dS spacetime, they both tend to $M_{\rm Pl}^2/H^2$. If our results could be extrapolated to 4D, we would then expect a Page-like curve in which the fine-grained entropy saturates to a constant $S \approx S_{\rm dS} = M_{\rm Pl}^2/H^2$, rather than decreasing, as in Fig.~\ref{fig:Page}. Nevertheless, a direct 4D calculation looks unfeasible, since the entanglement entropy of a CFT is much less known in this case, and the CFT gravitates. 

Let us comment on point 2 above. The island found in this work is not on the same Cauchy slice as the region $\mathcal{R}$ (since it is time-like separated from it) and for this reason an interpretation of the island formula in terms of the entanglement between $\mathcal{R}$ and $\mathcal{I}$ looks involved. 
At first sight, such interpretation appears to make sense only if $\mathcal{R}$ and $\mathcal{I}$ are independent subsystems, being part of the same Cauchy slice.
 Therefore, it would be interesting to investigate further the quantum-information meaning of time-like separated islands, such as the one that we found. 
 
 Here, we just give some intuitive considerations. As shown in Fig.~\ref{fig:sketchisland}, even though $\mathcal{I}$ and $\mathcal{R}$ are not on the same Cauchy slice, they ``almost'' are, although in an unusual sense: all light-like trajectories (the ones relevant for the CFT) may intersect either $\mathcal{I}$ and $\mathcal{R}$, but not both. In this sense, for the CFT they are independent systems. The key point is then, as shown in Fig.~\ref{fig:sketchisland}, many entangled pairs naively contributing to the entropy of $\mathcal{R}$ are indeed purified\footnote{More precisely, for a CFT the contributions of the left- and right-movers add up, as shown for instance by the form of~\eqref{eq:one_interval} and \eqref{eq:two_intervals}.
Then, many entangled right-moving modes around the right endpoint of $\mathcal{R}$ are purified by $\mathcal{I}$, as well as the left-moving modes close to the left endpoint. Instead, the left-movers around the right endpoint, as well as the right-movers around the left endpoint, are not purified, but in the limit $\epsilon \to 0$ this is sufficient to formally suppress completely the entanglement entropy. This is the same reason why the entropy of a single light-like interval tends to zero.} by $\mathcal{I}$, and this is the extremal surface achieving that. This is why the island calculation corrects the naive semiclassical expectation.

We do not know which saddles should be included in the Euclidean gravitational path integral, and what constitutes a legitimate extremum for the island formula, in particular whether a local maximum, such as our result, is acceptable. In the BH context the only legitimate islands are thought to be maximin saddle-points (maximal along time, minimal along space). In the dS case this is not known, especially for a time-like separated island such as ours. We can only check that the obtained entropy is free of obvious pathologies. In a context similar to ours, the authors of \cite{Chen:2020tes} found a minimax island (minimal along time, maximal along space), but then argued that it violates the strong-subadditivity condition, a basic criterion for entropies. In Appendix~\ref{app:SSA}, we review their argument and show that our island does not lead to a similar paradox. 
Finally, notice that the entropy \eqref{eq:Ssemi_island} is real, as it should be\footnote{The authors of~\cite{Chen:2020tes} obtain a complex entropy, but point out that their imaginary part cancels between bra and ket sheets.}, thanks to the fact that both $BC$ and $BD$ are time-like, a property that in turn implies the quasi-Cauchy property discussed above. 
 
To conclude, if our results could be extrapolated to 4D, they would suggest that the entropy bound on the duration of dS is actually absent. This is not the only obstruction to the possibility of a long (but possibly finite) stage of dS evolution \cite{Dubovsky:2008rf,Dvali:2014gua,Obied:2018sgi,Garg:2018reu,Dvali:2021kxt}, and thus it would certainly be interesting to investigate whether these other arguments could be affected, in a similar way, by the considerations presented here. For instance, the argument of~\cite{Ooguri:2018wrx} to corroborate the dS conjecture of~\cite{Obied:2018sgi,Garg:2018reu}, is precisely based on the entropy bound in the presence of many degrees of freedom. It would be interesting, albeit challenging, to study the consequences of our findings for these matters in future work.

\section*{Acknowledgments}
We thank 
Raffaele D'Agnolo and Tevong You for discussions and comments, and especially Gian Giudice and Matthew McCullough for, in addition, having drawn originally our attention to this problem. We are particularly grateful to Victor Gorbenko for having pointed out to us the correct implementation of the cutoff prescription in the island region.


\appendix

\section{No strong-subadditivity paradox}\label{app:SSA}

In a similar context to ours, the authors of~\cite{Chen:2020tes} found an island under certain assumptions, analogous to our subdominant extremum in Fig.~\ref{fig:close_past} and argued that this is pathological, because it violates the strong-subadditivity (SSA) condition
\beq
S(\mathcal{A} \cup \mathcal{B}) + S(\mathcal{A} \cup \mathcal{C}) \geq S(\mathcal{A}) + S(\mathcal{A} \cup \mathcal{B} \cup \mathcal{C}) \,,
\eeq
which is a basic requirement for an entropy. Their argument goes as follows. One introduces a \textit{probe} CFT with a small central charge $c_p \ll c$, which does not affect the location of the island. In the flat region the two CFTs can be considered as distinct systems. Then, since even in their case $x_*$ is a decreasing function of $c$, one can choose a region $\mathcal{R}$ such that the island dominates for $c$ but not for $c_p$. Then, the SSA can be calculated for the three distinct systems $\mathcal{R}_c$, $\mathcal{R}_{c_p}$, $\overline{\mathcal{R}}_{c_p}$, obtaining:
\beq
S(\mathcal{R}_{c} \cup \mathcal{R}_{c_p}) + S(\overline{\mathcal{R}}_{c_p}\cup \mathcal{R}_{c_p}) - S(\mathcal{R}_{c} \cup \overline{\mathcal{R}}_{c_p}\cup \mathcal{R}_{c_p}) - S(\mathcal{R}_{c_p}) \simeq - \frac{2 c_p}{3} \log \frac{l_{\mathcal{R}} c }{\phi_r} \ll -1 \,
\eeq
so that the SSA condition is violated by a large universal amount, logarithmic in the size of $\mathcal{R}$. 

We can repeat an analogous argument in our case. The two CFTs are not interacting, so 
\beq
S(\overline{\mathcal{R}}_{c_p}\cup \mathcal{R}_{c_p}) = 0 \,.
\eeq
For the probe CFT the non-island entropy is dominating, therefore:
\beq
S(\mathcal{R}_{c_p})  \simeq \frac{c_p}{3} \log \frac{2 x e^{H t_0}}{\epsilon} \,.
\eeq
The fine-grained entropy of an interval for both CFTs is given by
\beq
S(\mathcal{R}_{c} \cup \mathcal{R}_{c_p})  \simeq 2\phi_0 - \phi_r H x + \frac{c+c_p}{3}  \bigg(\log \frac{2 e^{H t_0} (c+c_p)}{9 \phi_r H^2\epsilon_{\rm RG}  \epsilon } - 1\bigg) \,.
\eeq
Finally, the entropy of the three subsystems is obtained by neglecting the effect of the probe CFT on the location of the island. However, the cut in the replica manifold given by the island affects all fields, including the probe CFT. The total entropy is thus given by the two-interval result for $c$, plus the $c_p$ single-interval result on the island:
\be
S(\mathcal{R}_{c} \cup \overline{\mathcal{R}}_{c_p} \cup \mathcal{R}_{c_p}) \simeq 2\phi_0 - \phi_r H x + \frac{c}{3}  \bigg(\log \frac{2 e^{H t_0} c}{9 \phi_r H^2\epsilon_{\rm RG}  \epsilon } - 1\bigg) + \frac{c_p}{3} \log \frac{c}{3 H^2 \epsilon_{\rm RG} \phi_r  x} \,.
\ee
All in all, we obtain:
\beq
S(\mathcal{R}_{c} \cup \mathcal{R}_{c_p}) + S(\overline{\mathcal{R}}_{c_p}\cup \mathcal{R}_{c_p}) - S(\mathcal{R}_{c} \cup \overline{\mathcal{R}}_{c_p}\cup \mathcal{R}_{c_p}) - S(\mathcal{R}_{c_p}) \simeq \frac{c_p}{3} \log \frac{1}{3} \approx 0 \,,
\eeq
which vanishes within the approximations of our formulas (which in particular disregard the non-logarithmically-enhanced part of the entropy of a CFT). Therefore, our results do not violate the SSA condition (at least in this setup), but rather saturate it.


\bibliography{refs}
\bibliographystyle{utphys}


\end{document}